Piezo-Ferroelectrics: Pressure-induced phase transitions in Pb(Zr$_{1-x}$Ti$_x$)O$_3$ (PZT)

**Jérôme Rouquette[1,2], Julien Haines[1], Véronique Bornand[1] and Philippe Papet[1]**

[1]Laboratoire de Physico-Chimie de la Matière Condensée, UMR CNRS 5617, Université Montpellier II, Place Eugène Bataillon, cc003, 34095 Montpellier cedex 5, France.
[2]Bayerisches Geoinstitut, Universität Bayreuth, D-95440 Bayreuth, Germany.

## Abstract

Piezoelectric PZT solid solutions were studied as a function of pressure and temperature. Moderate pressure is found to induce phase transitions to monoclinic phases (Cm,Cc) for Ti-rich PZT, which are the same ferroelectric phases responsible for the high piezoelectric properties of the MPB. The following transformation sequence P4mm → Cm → Cc → F1 → F$\bar{1}$ was identified upon increasing pressure. Compression behaviour is dominated by: (1) a rotation and a reduction of the spontaneous polarization and (2) octahedral tilting leading to unit cell doubling. Future studies as a function of pressure, temperature and electric field on single crystals of heterovalent relaxor ferroelectric isotypes of PZT are proposed in order to to better understand these potentially technologically-important piezo- ferroelectric materials.

# Introduction

Lead zirconate titanate solid solutions Pb(Zr$_{1-x}$Ti$_x$)O$_3$ (PZT) represent one of the most important classes of piezoelectric materials because of their excellent properties. These materials are used in different systems such as piezoceramics (piezoelectric transducers and actuators) and thin films (MEMS system or high-frequency components). The highest piezoelectric constants and electromechanical coupling coefficients have been obtained for compositions near the morphotropic phase boundary (MPB) between rhombohedral (*R3m*) and tetragonal (*P4mm*) phases (x ≈ 0.48)[1]. Recently the MPB has attracted particular attention due to the discovery of a monoclinic phase (*Cm*) in this composition range at temperatures below 300K[2]. In this monoclinic form, the polarization can be modulated for example as a function of an external applied electric field between the rhombohedral and tetragonal directions based on a polarization rotation mechanism thus providing a possible explanation for this high piezoelectric response[3]. In addition, a doubling of this monoclinic unit cell was reported at low temperature (T ≤ 210 K) for $PbZr_{0.52}Ti_{0.48}O_3$ due to antiferrodistortive tilting of the $(Zr,Ti)O_6$ octahedra based on the appearance of superlattice reflections in electron[4] and neutron diffraction[5] data. This complex phase diagram around the MPB does not take into account further local structural distortions arising from the static disorder of the zirconium and titanium ion displacements[6,7] which are, in addition, at the origin of the well-known ferroelectric domain structure. Furthermore, recent experiments based on electron diffraction[6] and neutron diffraction studies[8], Raman scattering results[9], as well as theoretical calculations[10] have indicated the existence of an intrinsic short range dynamic disorder over nearly the entire PZT solid solution predominantly due to off-center lead displacements[8,11]. Therefore, the boundaries between the different phases, in particularly close to the MPB are still subject to debate.

Due to their high polarizability, ferroelectric materials exhibit large dielectric constants and excellent piezoelectric properties. These characteristics are extremely dependent on three variables (for any given composition): the applied electric field (E), the mechanical stress ($\sigma_{ij}$) and the temperature (T)[12]. In addition to the ongoing investigation of the relatively well-studied, highly complex, composition-temperature phase diagram of PZT, the effect of hydrostatic pressure ($\sigma_{ij}$ = P for i = j) is therefore of great importance in order to better understand the dependence of both structural and dielectric properties on the two fundamental variables P and T for such a ferroelectric system. In the present review, recent investigations of PZT as a function of pressure and temperature by neutron and x-ray diffraction, Raman spectroscopy and dielectric measurements to characterize the multi-scale behaviour of this material at the nano-micro-macro-scopic range are presented.

## 2. Study of morphotropic $PbZr_{0.52}Ti_{0.48}O_3$

Atmospheric pressure neutron diffraction data as a function of the temperature were consistent with the presence of the ambient-temperature (*Cm*) and low-temperature (*Cc*) monoclinic forms[13] in $PbZr_{0.52}Ti_{0.48}O_3$. Interestingly, the associated temperature phase transition sequence observed by diffraction[4,5] and dielectric measurements[14,15] could also clearly be distinguished by resonance Raman spectroscopy[16]. In the literature, Raman measurements reported for composition close to the MPB showed only minor spectral changes in the temperature-composition range consistent with the static and dynamic disorder associated with the PZT solid solution[17,18,19,20], Fig. 1a. Using an excitation line of 647.1 nm with an energy close to a self-trapped level exciton energy deficient complex (Ti$_{Ti'}$-V$_O^{..}$)[21,22,23] of PZT powder, a resonance effect was induced. In order to illustrate this difference in behaviour, Raman spectra obtained as a function of temperature with both 647.1 nm and 514.5 nm excitation lines are represented on Fig. 1.

Resonance Raman spectroscopy clearly allows the structural changes of PZT to be identified as a function of P and T[13,16,24,25,26].

Early high pressure studies of the morphotropic composition PbZr$_{0.52}$Ti$_{0.48}$O$_3$ were restricted to the determination of the phase boundary between the ferroelectric tetragonal and cubic paraelectric forms at high temperatures for pressures up to 0.8 GPa[27,28].

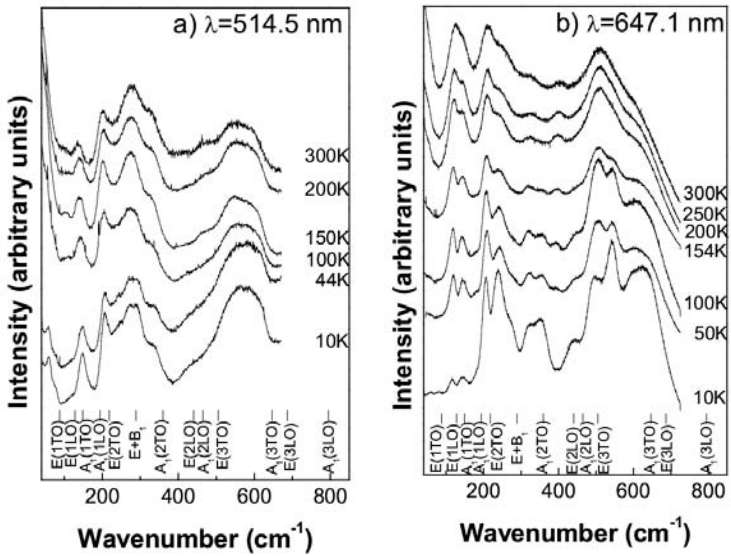

**Figure 1.** Raman spectra obtained between 10 and 300 K with a) λ$_{laser}$ = 514.5, and b) λlaser = 647.1 nm. Positions of the Raman peaks at room temperature for PbTiO$_3$ are indicated[29].

Neutron diffraction data indicate that at close to room temperature, the cubic form is not obtained at high pressure but instead a series of low-symmetry phases are obtained. Superlattice reflections, arising from a doubling of the elementary pseudo-cubic perovskite unit cell induced by antiferrodistortive tilting of the $(Zr,Ti)O_6$ octahedra. are found to appear at low pressure (P $\geq$ 1.1 GPa) and are characteristic of the monoclinic *Cm-Cc* phase transition, the same transformation that is observed at low temperature and ambient pressure[5]. Note the importance of the use of neutron diffraction, which enables to distinguish superlattice reflections to be detected. These would be extremely weak in x-ray diffraction. Neutron diffraction data obtained at 288 K and 3.8 GPa are consistent with a triclinic structure, space group *F*1 (Fig. 2)[30]. The superlattice reflections further increase in intensity in this phase.

      The structural data obtained from neutron diffraction experiments can be used to calculate the spontaneous polarization ($P_s$) and the polarization rotation angle (θ) with respect to the pseudo-tetragonal z-axis in the monoclinic phase (Fig. 3).

      Additionally, the high-pressure distorted perovskite *F*1 phase exhibits a second rotation angle φ (calculated from $P_{s(x)}$ and $P_{s(y)}$) to describe the orientation of the spontaneous polarization. The spontaneous polarization calculated from the structural data at 3.8 GPa and 288 K is 17 μC/cm$^2$ with an in plane ($1\bar{1}0$) component of rotation angle (θ) of 30°. With increasing pressure, a transformation to a pseudo-cubic paraelectric phase ($F\bar{1}$) has been identified.

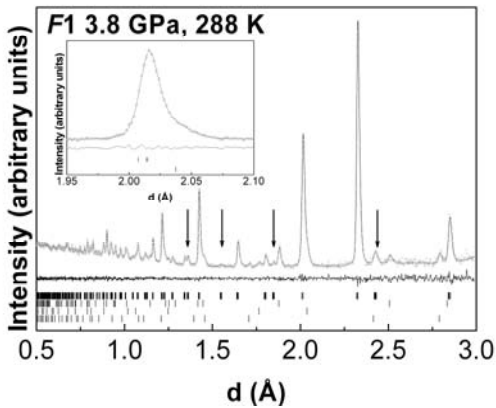

**Figure 2.** Experimental and calculated profiles from Rietveld refinements of the triclinic *F*1 phase of $PbZr_{0.52}Ti_{0.48}O_3$ at 288 K and 3.8 GPa. The difference profile is on the same scale. Vertical ticks indicate, from the top down, the calculated positions of the *F*1 phase, tungsten carbide and nickel from the anvils of the pressure cell and lead. Arrows indicate the strongest superlattice reflections.

　　　　　In this centrosymmetric structure, Fig. 4, the Zr, Ti cations are located on inversion centers and the constituent octahedra are tilted about directions parallel to **a**, **b** and **c**.
Such a space group is consitent with the persistence of the observed high-pressure Raman spectra in the paraelectric state[13,16,24,25,26]. In contrast to ambient-pressure paraelectric cubic $Pm\bar{3}m$ structure, the first-order Raman signal is not forbidden by group theory for the $F\bar{1}$ phase.
　　　　　The results obtained enable a *P-T* phase diagram to be construct for PbZr$_{0.52}$Ti$_{0.48}$O$_3$, Fig. 5. In this diagram, data obtained from Raman spectroscopy, dielectric measurements and x-ray diffraction are also reported. Additionally, the spontaneous polarization vector calculated from neutron diffraction studies are represented for each *P-T* point. The polarization rotation mechanism and the associated reduction in polarization as a function of *P* and *T* are clearly visible.

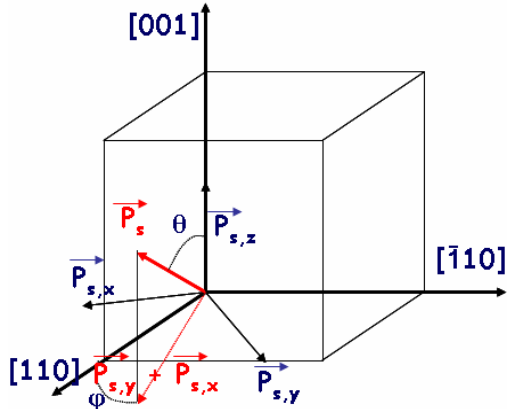

**Figure 3.** Spontaneous polarization $P_s$ (three components: $P_{s(x)}$, $P_{s(y)}$ and $P_{s(z)}$) with the two rotation angles (θ, φ) in triclinic $F1$ symmetry. In monoclinic phases (space group *Cm*, *Cc*), $P_s$ has two components ($P_{s(x)} = P_{s(y)}$ and $P_{s(z)}$) with one rotation angle (θ).

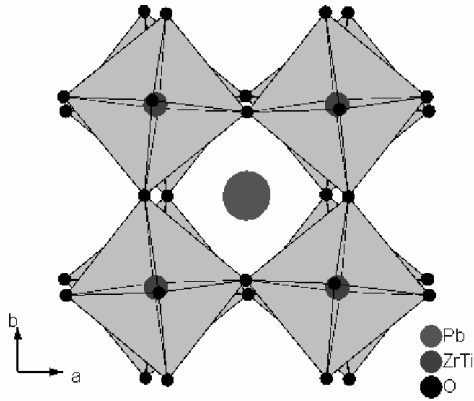

**Figure 4.** Crystal structure of the triclinic $F\bar{1}$ phase of PbZr$_{0.52}$Ti$_{0.48}$O$_3$.

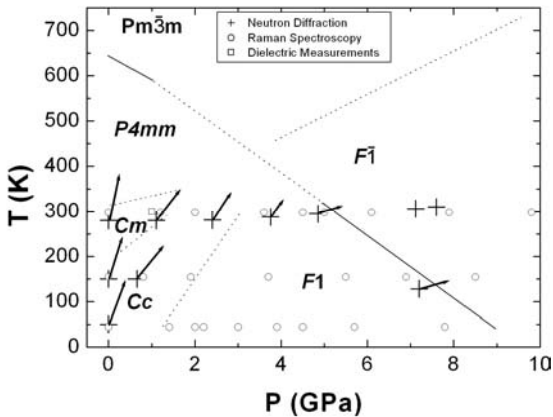

**Figure 5.** P-T phase diagram of PbZr0.52Ti0.48O3. The solid lines represent phase boundaries obtained from X-ray diffraction, Raman spectroscopy and dielectric measurements; Dashed lines correspond to proposed P-T boundaries.

## 3  Study of Ti-rich PZT

The data obtained by neutron diffraction on PbZr$_{0.40}$Ti$_{0.60}$O$_3$ (which is tetragonal under ambient conditions) indicated the presence of the monoclinic structures *Cm* and *Cc* respectively at 2.7 and 4.1 GPa, Fig. 6a and 6b, the same ferroelectric forms which are proposed to be responsible for the high piezoelectric properties characteristic of the MPB composition PbZr$_{0.52}$Ti$_{0.48}$O$_3$ at ambient pressure[31]. This means that hydrostatic pressure can tune the MPB and its associated exceptional physical properties by stabilizing the monoclinic phases to include compositions that do not normally lie on the MPB. Both phase transitions were also clearly characterized from spectral changes (frequency, intensity and damping) in Raman spectroscopic data and from the changes in compressibility in an x-ray diffraction study. Additionally, the *P4mm-Cm* transformation was also identified from the change in slope of the inverse of the dielectric constant ($1/\varepsilon'_r$) as a function of pressure. The latter *Cm-Cc* phase transition can directly be detected by neutron diffraction from the appearance of superlattice reflections, Fig. 6, indicating a doubling of the unit cell due to octahedral tilting.

Fig. 7 represents the 111 and 200 pseudocubic reflections of PbZr$_{0.40}$Ti$_{0.60}$O$_3$ at 9.6 GPa (highest pressure reached from neutron diffraction experiments) obtained from Rietveld refinements using either the rhombohedral *R3c* or triclinic *F*1 space group[32]. Note that the *Cc* structure was also used as a starting model. Out of these three structural models, the worse refinement was obtained using the rhombohedral structure (*Rwp* = 5.7%, *Rp* = 8.7%). The difference from experimental and calculated profiles can clearly be seen in Fig. 7. The 111 reflection should be split for a *R3c* structure a relative intensity ratio of 0.9. This intensity ratio does not correspond to the observed data. Similarly, the 200 reflection is not split in rhombohedral symmetry, which thus cannot reproduce the experimental data. Consequently the space group *R3c* is not observed at high pressure for these PZT solid solutions.

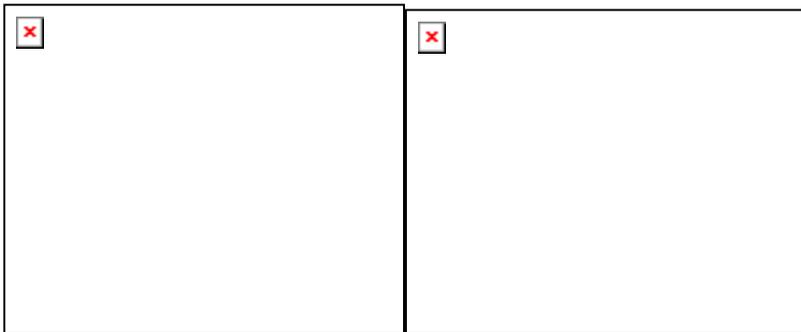

**Figure 6.** Experimental and calculated profiles from Rietveld refinements of (a) the monoclinic *Cm* phase of PbZr$_{0.40}$Ti$_{0.60}$O$_3$ at 2.7 GPa and (b) the monoclinic *Cc* phase of PbZr$_{0.40}$Ti$_{0.60}$O$_3$ at 4.1 GPa. The difference profile is on the same scale. Vertical ticks indicate the calculated positions of the PZT phase (top) and of tungsten carbide (middle) and nickel (bottom) from the anvils of the pressure cell.

The strongest superlattice reflection is shown in the inset. Arrows indicate the most prominent superlattice reflections.

The best refinement is obtained using the triclinic model (*Rwp* = 4.9%, *Rp* = 7.3%). In such a structure, rotations angles $\omega_a$ and $\omega_b$ which are identical in *Cc* symmetry become different, Table 1, and the spontaneous polarization is described with a second rotation angle φ. Use of a Paris-Edinburgh high-pressure cell with tungsten carbide cells limits this work to pressures of about 10 GPa. The paraelectric phase $F\bar{1}$ observed for PbZr$_{0.52}$Ti$_{0.48}$O$_3$ was not identified in this work for PbZr$_{0.40}$Ti$_{0.60}$O$_3$. Extrapolation of $P_s \rightarrow 0$ using an arbitrary second order polynomial can give a good insight of the ferroelectric-paraelectric pressure transformation. The corresponding value is about 10-11 GPa in accordance with results obtained from x-ray diffraction and Raman spectroscopy measurements. By analogy with PbZr$_{0.52}$Ti$_{0.48}$O$_3$, the structural transition $F1 \rightarrow F\bar{1}$ is proposed for this ferroelectric-paraelectric transformation.

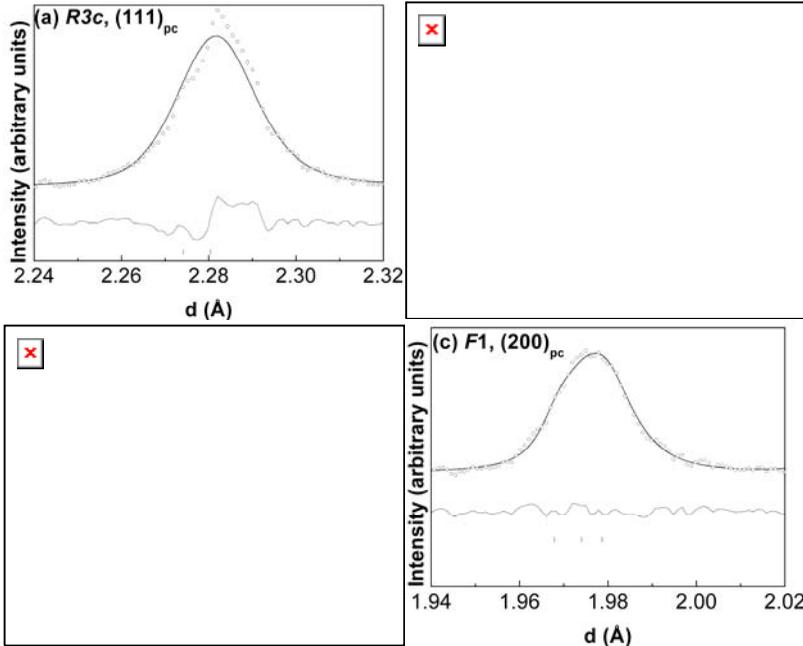

**Figure 7.** Experimental and calculated profiles from Rietveld refinements of (a) the (111) and (b) (200) pseudocubic reflections *R3c* phase of PbZr$_{0.40}$Ti$_{0.60}$O$_3$ at 9.6 GPa. (c) (111) and (d) (200) pseudocubic reflectionc *F*1 phase of PbZr$_{0.40}$Ti$_{0.60}$O$_3$ at 9.6 GPa. The difference profile is on the same scale. Vertical ticks indicate the calculated positions of the PZT phase.

The in-plane rotation angle of the spontaneous polarization ($\theta$) increases from 0° in the *P4mm* structure to 76° in the *F*1 form. This value is close to the pseudocubic [112] direction previously observed for the polarization vector in the recent study of the *F*1 structure in PLZT ($\theta=70.5°$)[33], indicating that the rotation of the in-plane component of $P_s$ principally occurs between the [001] and the [112] directions. This high-pressure rotation of $P_s$ is found to be a general feature of PZT solid solutions. Finally, the high-pressure compression behavior is dominated by octahedral tilting. This compression mechanism appears at the *Cm-Cc* phase transition. The *Cc* phase is characterized by two distinct rotation angles $\omega_a$ and $\omega_c$. These angles increase on compression and can be followed by the increase in intensity of the superlattice reflections. At the *Cc-F*1 phase transition, $\omega_a$ and $\omega_b$ become distinct as expected in triclinic symmetry.

| P (GPA) | T (K) | $P_S$ ($\mu C/CM^2$) | $\theta$ (°) | $\varphi$ (°) | $\omega_A$ (°) | $\omega_B$ (°) | $\omega_C$ (°) |
|---|---|---|---|---|---|---|---|
| 0.0001 | 280 | 36.8 | 0 | 0 | 0 | 0 | 0 |
| 2.7 | 295 | 32.6 | 4.3 | 0 | 0 | 0 | 0 |
| 4.1 | 295 | 30 | 42 | 0 | 0.8 | 0.8 | 2.6 |
| 6.6 | 295 | 23 | 48 | 0 | 2 | 2 | 4.4 |
| 9.6 | 295 | 8 | 76 | 24 | 3 | 3.1 | 4.3 |

**Table 1.** Spontaneous polarization $P_s$ and its associated rotation angles ($\theta,\varphi$), octahedral rotation angles ($\omega_a,\omega_b,\omega_c$) as a function of P and T.

The titanium-rich PZT solid solution PbZr$_{0.20}$Ti$_{0.80}$O$_3$ was studied under high pressure by X-ray diffraction and Raman spectroscopy[26]. Four high-pressure phase transitions were observed as in the case of PbZr$_{0.40}$Ti$_{0.60}$O$_3$. Based on the similarity of the Raman spectra and the P-V data obtained by x-ray diffraction for both compositions, it was proposed that the same phase transition sequence was observed (albeit at higher pressures), including monoclinic (*Cm*, *Cc*) and triclinic forms (*F*1 and $F\bar{1}$). In particular, high pressure Raman spectra obtained are consistent with a monoclinic (*Cc*) - triclinic (*F*1) phase transition involving a soft mode at 9 GPa, in agreement with the change in compressibility detected by X-ray diffraction. The strong Raman spectra observed in the paraelectric phase is also in agreement with triclinic $F\bar{1}$ symmetry.

Dielectric measurements of $Pb(Zr_{1-x}Ti_x)O_3$ for x = 0.48, 0.50, 0.52, 0.56 and 0.60 were studied from ambient pressure up to 2 GPa[31,34], Fig. 8. Two distinct types of behaviour can be observed: 1) the dielectric constant of $Pb(Zr_{0.52}Ti_{0.48})O_3$ is characterized by a hysteresis loop with a change in slope of $\varepsilon'_r$ consistent with a phase transition that does not have the same value on compression and decompression and which are represented by arrows on Fig. 8; 2) $\varepsilon'_r$ of PZT for x = 0.52, 0.56 and 0.60 have the same trend on compression and decompression. From the change in slope of the inverse of the dielectric constant ($1/\varepsilon'_r$) as a function of pressure, a transformation can also be detected. Effectively, the dielectric constant is known to diverge at the ferroelectric-paraelectric phase transition but also at a ferroelectric-ferroelectric one as a function of temperature, electric field and pressure[12]. These two kind of tendencies can be linked to two different phase transitions that were previously observed by neutron diffraction, *Cm-Cc* for $Pb(Zr_{0.52}Ti_{0.48})O_3$, *P4mm-Cm* for the rich-Ti compositions. The former could be ideally detected around the same transformation pressure by a change in volume from the compressibility of the material in x-ray diffraction data, from the appearance of new bands in Raman spectra associated to the doubling of the unit cell or by superlattice reflections in neutron diffraction. The latter transformation was characterized by a change in slope of the volume under pressure, by spectral changes in Raman scattering or by fine features like the line shape in neutron diffractogram. Interestingly, $Pb(Zr_{0.50}Ti_{0.50})O_3$ exhibits both phase transition from dielectric measurements, Fig. 8, respectively around 0.8 and 1.6 GPa. This is clearly consistent from our structural data which showed that this composition was tetragonal, space group *P4mm*, at ambient pressure and temperature[35].

The *P4mm* and *Cm* phases of PZT are known to exhibit both dynamic and static disorder from Raman spectroscopy[9], neutron diffraction[8] and electronic diffraction[6,7]. The reproducibility of dielectric measurements in both compression and decompression data associated to the same value of the transformation pressure imply a continuous structural phase transition. This continuous behaviour was already observed by x-ray diffraction (compressibility modulus), neutron diffraction (weak value of the rotation angle of $P_s$ for $Pb(Zr_{0.40}Ti_{0.60})O_3$ at 2.7 GPa) or even Raman spectroscopy ($\frac{\partial \nu}{\partial P}$). In this way, this *P4mm-Cm* transformation is closer to a second order rather than a first order phase transition. From group theory[36], this phase transition has therefore both a ferroelectric and a ferroelastic order parameter, respectively the rotation of the spontaneous polarization in the $(1\bar{1}0)$ plane and the primary order parameter: the spontaneous deformation e$_4$ (2e23) =-e$_5$ (2e13).

The *Cm-Cc* transformation is purely first order consistent with a change in volume from x-ray diffraction data due to the absence of the relevant compression mechanism "octahedral tilting" in *Cm* space group. Such a first order transformation must generate hysteresis phenomenon (thermodynamic equilibrium) and kinetics behaviour which could explain the different transition values obtained by neutron diffraction and dielectric measurements. This event was also clearly observed from the Raman spectral changes as a function of pressure ($\frac{\partial \nu}{\partial P}$) which were associated to a coexistence of phases. This could also explain electronic diffraction data from Noheda et al. which proposed the conservation of the *Cm* phase down to 10 K with a small amount of *Cc* proportion (30%)[37]. These findings could be explained by a monoclinic *Cm-Cc* solid state transformation associated to low kinetics. From group theory this phase transition is non-ferroïc, its order parameter is therefore the tilting in antiphase of the oxygen octahedral.

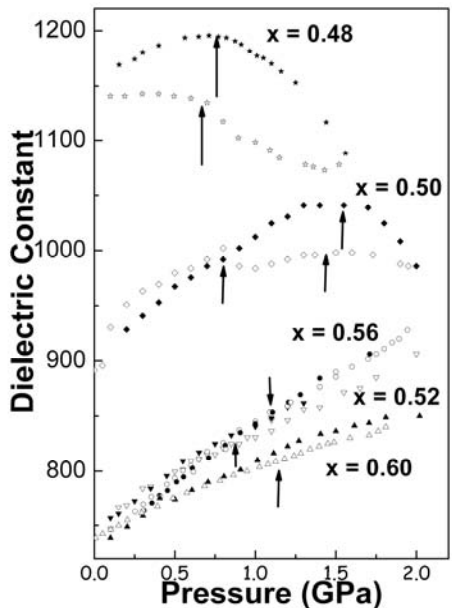

**Figure 8.** Dielectric constant of PbZr$_{1-x}$Ti$_x$O$_3$ measured at 100 kHz versus pressure for selected compositions (x). Pressure-cycling was performed for each composition. Solid and empty symbols correspond to values obtained on compression and decompression, respectively. Both *P4mm-Cm* and *Cm-Cc* phase transition are represented with arrows.

## 4. Conclusion and future developments

The effects of pressure and temperature on the most commonly used piezoelectric perovskite solid solution PZT are reviewed in the present work. Based on x-ray and neutron diffraction, dielectric measurements and Raman spectroscopy it is shown that high pressure can tune the morphotropic phase boundary (associated with the exceptional piezoelectric properties) in the composition-pressure plane to include compositions that do not normally lie in the MPB at ambient pressure. Thus, by controlling the appropriate composition and stress state of materials in ceramics and thin films, optimal piezoelectric properties may be obtained for technological applications. This corresponds to a potentially direct application of the pressure parameter to high-performances ferroelectric compounds. Upon increasing pressure, a sequence of low-symmetry structures are identified (*Cm*, *Cc*, $F1$, $F\bar{1}$). The high pressure behaviour is consistent with a reduction and a rotation of the spontaneous polarization and the onset of octahedral tilting leading to unit cell doubling. The results obtained have enabled us to construct a pressure-temperature, Fig. 4, and a pressure-composition phase diagram of $PbZr_{0.52}Ti_{0.48}O_3$, Fig. 8.

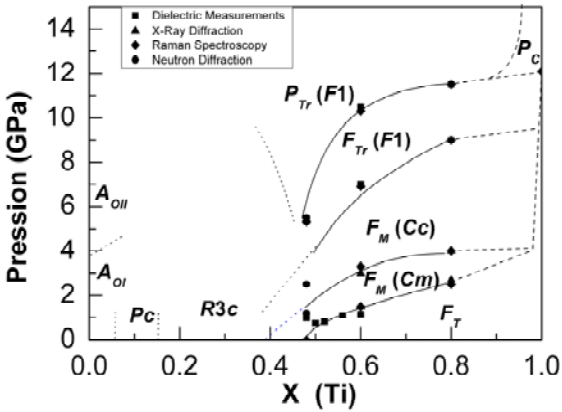

**Figure 8.** Pressure-composition phase diagram of $Pb(Zr_{1-x}Ti_x)O_3$. Solid lines are boundaries from the present study. Dashed lines are extrapolation to lead titanate end member with the paraelectric transition to $Pm\bar{3}m$ at 12 GPa from Raman data[38]. Note that recent first principles calculations reported different high pressure

phase transition sequence for PbTiO$_3$[39]. Dot lines correspond to boundaries proposed from rich-Zr composition. Note that our first results show that *R3m* is not the correct phase even at ambient pressure. *Pc* and $A_{OI}$ ~ *Pbam* were proposed by Woodward et al.[7]. $A_{OI}$ - $A_{OII}$ orthorhombic phase transition is proposed based on high pressure x-ray diffraction and dielectric measurements[40].

Preliminary results support the extrapolation of the *Cm-Cc* phase boundary towards rich-Zr compositions[41]. These new data tend to indicate that the rhombohedral *R3m* phase may not be stable even at atmospheric pressure, as previously suggested by Pandey et al., and that *Cm* may be the stable phase[42]. This would imply direct group-subgroup relationships at ambient conditions from the *P4mm* tetragonal space group of PbTiO$_3$ end-member to the *Cm-Cc* monoclinic forms of the MPB and then to the *R3c* supergroup contrary to the originally proposed *P4mm-R3m-R3c* series of transformations. The boundary with the paraelectric phase is proposed to have a positive slope from the MPB to rich-Zr compositions based on these first preliminary results. This interesting finding implies the existence at least of another phase at high pressure. These features along with the probable complex phase transition sequence for compositions close to lead zirconate make these further investigations particularly necessary and challenging.

As previously mentioned, ferroelectric properties are dependent to the three parameters: stress-electric field-temperature. The present work shows the importance to undertake coupled investigations as a function of P and T in order to determine the influence of two of the three variables on which ferroelectric properties depend. In the future, the next step will be to extend these studies to well-known relaxors-ferroelectric solid solutions (1-x)Pb(Zn$_{1/3}$Nb$_{2/3}$)O$_3$-xPbTiO$_3$ + PMN-PT which are heterovalent isotypes of PZT, exhibit a MPB region and shows giant piezoelectric response (superior to PZT) in single crystal form (contrary to PZT). The single crystal form also provides three dimensional information and enables the materials to be polarized in the desired orientation. In this way, it would be possible to characterize these materials as a function of the three fundamental variables, *P*, *T*, and $\vec{E}$, to have an overall view and a better understanding of this particular family of materials.